\documentclass[useAMS,usenatbib,fleqn]{mn2e}
\usepackage[latin1]{inputenc}
\usepackage{amsmath}
\usepackage{amsfonts}
\usepackage{amssymb}
\usepackage{setspace}
\usepackage{epsfig}
\usepackage{lscape,graphicx}
\usepackage{natbib}
\usepackage{textcomp}
\usepackage{epstopdf}
\usepackage{graphicx}
\usepackage{caption}
\usepackage{float}
\usepackage{aas_macros}

\def\lsim{\mathrel{\lower0.6ex\hbox{$\buildrel {\textstyle <}
 \over {\scriptstyle \sim}$}}}
\def\gsim{\mathrel{\lower0.6ex\hbox{$\buildrel {\textstyle >}
 \over {\scriptstyle \sim}$}}}

\begin{document}

\title[Galaxy Properties and the Cosmic Web]{Galaxy properties and the cosmic web in simulations} 

\author[Metuki et al.] 
{\parbox[t]\textwidth{ Ofer Metuki$^1$, Noam I. Libeskind$^{2}$, Yehuda Hoffman$^1$, Robert A. Crain$^{3}$, Tom Theuns$^{4,5}$} 
\vspace*{6pt} \\ 
 $^{1}$Racah Institute of Physics, Hebrew University, Jerusalem 91904, 
 Israel\\ 
 $^{2}$Leibniz-Institut f\"ur Astrophysik Potsdam (AIP), An der Sternwarte 16, D-14482 
 Potsdam, Germany\\
 $^{3}$Leiden Observatory, Leiden University, PO Box 9513, 2300 RA, Leiden, the Netherlands\\
$^{4}$Institute for Computational Cosmology, Department of Physics, University of Durham, South Road, Durham, DH1 3LE \\
$^{5}$Department of Physics, University of Antwerp, Campus Groenenborger, Groenenborgerlaan 171, B-2020 Antwerp, Belgium\\
}

\date{\today}  

\maketitle

\begin{abstract}  
 
We seek to understand the relationship between galaxy properties and their local environment, which calls for a proper formulation of the notion of environment. We analyse the GIMIC suite of cosmological hydrodynamical simulations within the framework of the cosmic web as formulated by Hoffman et al., focusing on properties of simulated DM haloes and luminous galaxies with respect to voids, sheets, filaments and knots - the four elements of the cosmic web. We find that the mass functions of haloes depend on environment, which drives other environmental dependence of galaxy formation. The web shapes the halo mass function, and through the strong dependence of the galaxy properties on the mass of their host haloes, it also shapes the galaxy-(web) environment dependence.

\end{abstract}


\section{Introduction}  
\label{sec:intro}

Galaxy formation is one of the main research foci in modern cosmology, and one of the most challenging. According to the standard cosmological paradigm, the initial density field is nearly perfectly homogeneous, peppered with small perturbations that cause structures to form in the Universe. These perturbations grow via gravitational instability, creating a filamentary network known as the cosmic web \citep{1996Natur.380..603B}, delineated by geometrically distinct components termed voids, sheets, filaments, and knots. The cosmic web is the most striking manifestation of gravitational collapse on the scales of a few megaparsecs and above, and has been known since the advent of numerical simulations \citep{1985ApJ...292..371D, 1987ApJ...313..505W} and observed in large sky surveys, such as the CfA redshift survey \citep{1989Sci...246..897G}.

On the sub-megaparsec scale, dark matter (DM) collapses into virialized structures called haloes. If a DM halo's potential well is deep enough, it may act as a site of galaxy formation, enabling the efficient cooling of gas and the formation of stars. That certain galaxy properties (such as luminosity, colour, morphology, etc.) correlate with the mass of their host halo is a fairly well established feature of galaxy formation \citep{2000MNRAS.318.1144P,2002ApJ...575..587B,2010ApJ...710..903M}. However, it is not clear what role environment plays in the process of galaxy formation. \cite{1980ApJ...236..351D} has shown that galaxy morphology correlates with the galaxy density in their neighbourhood. More recent studies have shown correlations between the environment and both galactic properties \citep[e.g., luminosities, surface brightnesses, colours, and profile shapes; see][]{2005ApJ...629..143B} as well as host halo properties \citep[e.g.  mass, spin parameter, shape, and mass assembly histories; see][]{2005ApJ...634...51A,2005MNRAS.363L..66G,2007ApJ...654...53M}.  \cite{2012MNRAS.424.1179B} have also shown a correlation for satellite galaxies. These studies suggest that although galaxy formation occurs deep within a halo's potential well,  the process may be significantly influenced by larger scale phenomena. However, these studies adopt different definitions of environment, and it is therefore necessary to scrutinize this definition. In observational studies, projected galactic number density is often used to define environment \citep[a good overview of such methods can be found in][]{2012MNRAS.419.2670M}. In simulations one may use additional measures such as the DM density.

Definitions of environment \citep[e.g.][]{2007A&A...474..315A, 2008MNRAS.383.1655S, 2010MNRAS.406.1609B, 2010MNRAS.409..156B} often do not employ kinematical data, since these are challenging if not impossible to directly measure in observational samples. However, within the past few years, a number of dynamical techniques have been developed for the analysis of simulations \citep[e.g.][]{2007MNRAS.375..489H,2009MNRAS.396.1815F,2012MNRAS.425.2049H,2013MNRAS.429.1286C}. \cite{2007MNRAS.375..489H} and \cite{2009MNRAS.396.1815F} employed the tidal tensor, defined as the Hessian of the potential, to identify the cosmic web in cosmological simulations. More recently, \cite{2012MNRAS.425.2049H} used the  velocity shear tensor, as an extension to the tidal tensor, to define the cosmic web. This so-called V-web is able the probe the non-linear regime deep into the sub-megaparsec scale. A growing body of literature \citep{2006MNRAS.370.1422A, 2007A&A...474..315A, 2011MNRAS.413.1973W, 2012MNRAS.427.3320C, 2012MNRAS.421L.137L, 2013MNRAS.428.2489L,2014MNRAS.441.2923C,2014MNRAS.441..646N} has employed these ``web finders'' to unveil properties of the DM haloes that populate the cosmos.

Many models of galaxy formation implicitly limit the effect the environment may have. For example, Halo Occupation Models \citep[HODs; e.g.][]{2007MNRAS.376..841V, 2006ApJ...647..201C} assume that galaxy properties are a function solely of the host halo's mass. More sophisticated semi-analytical models  \citep[SAMs; e.g.][]{1993MNRAS.264..201K, 1994MNRAS.271..781C, 2000MNRAS.319..168C} often assume galaxy properties are determined uniquely by a halo's merger history. Cosmological hydrodynamical simulations implicitly include the environmental effects on galaxy formation but progress along this avenue is hindered by their high computational cost and uncertainties in the subgrid physics.
Yet, progress has been made with the emergence of some recent simulations in which the limited dynamical range has been shifted to scales large enough so as to probe the environmental dependence of galaxy formation. This is the case of the Galaxies-Intergalactic Medium Interaction Calculation \citep[GIMIC;][]{2009MNRAS.399.1773C} suite of simulations, which are used here. 

In this paper we use the GIMIC suit of simulations to investigate how properties of galaxies differ across the cosmic web. The GIMIC suite of simulations has been used to study the properties of the hot, X-ray luminous circumgalactic medium \citep{2010arXiv1011.1906C, 2010MNRAS.407.1403C}, the orientation of satellite galaxies \citep{2011MNRAS.415.2607D}, stellar haloes \citep{2011MNRAS.416.2802F, 2012MNRAS.420.2245M}, star formation rates \citep{2012MNRAS.427..379M}, the origin of discs \citep{2012MNRAS.423.1544S}, and metal abundance in the circumgalactic medium \citep{2013MNRAS.432.3005C}, and is uniquely suited for our purposes as it combines a variety of environments (in the sense of the mean density) of moderate volume (which can be combined to statistically represent a much larger volume) with high resolution down to galactic scales, and (for the intermediate resolution) has been run all the way to redshift zero. The V-web algorithm \citep{2012MNRAS.425.2049H} is used to define the cosmic web of the GIMIC simulations.

The paper is organized as follows: \S\ref{sec:methods} describes the simulations and the cosmic web formalism and the results are presented in \S\ref{sec:results}. \S\ref{sec:disc} concludes the paper with a discussion of the results and their implications.

\section{Methods}
\label{sec:methods}

\begin{figure*}  
\begin{center} 
\includegraphics[width=13cm]{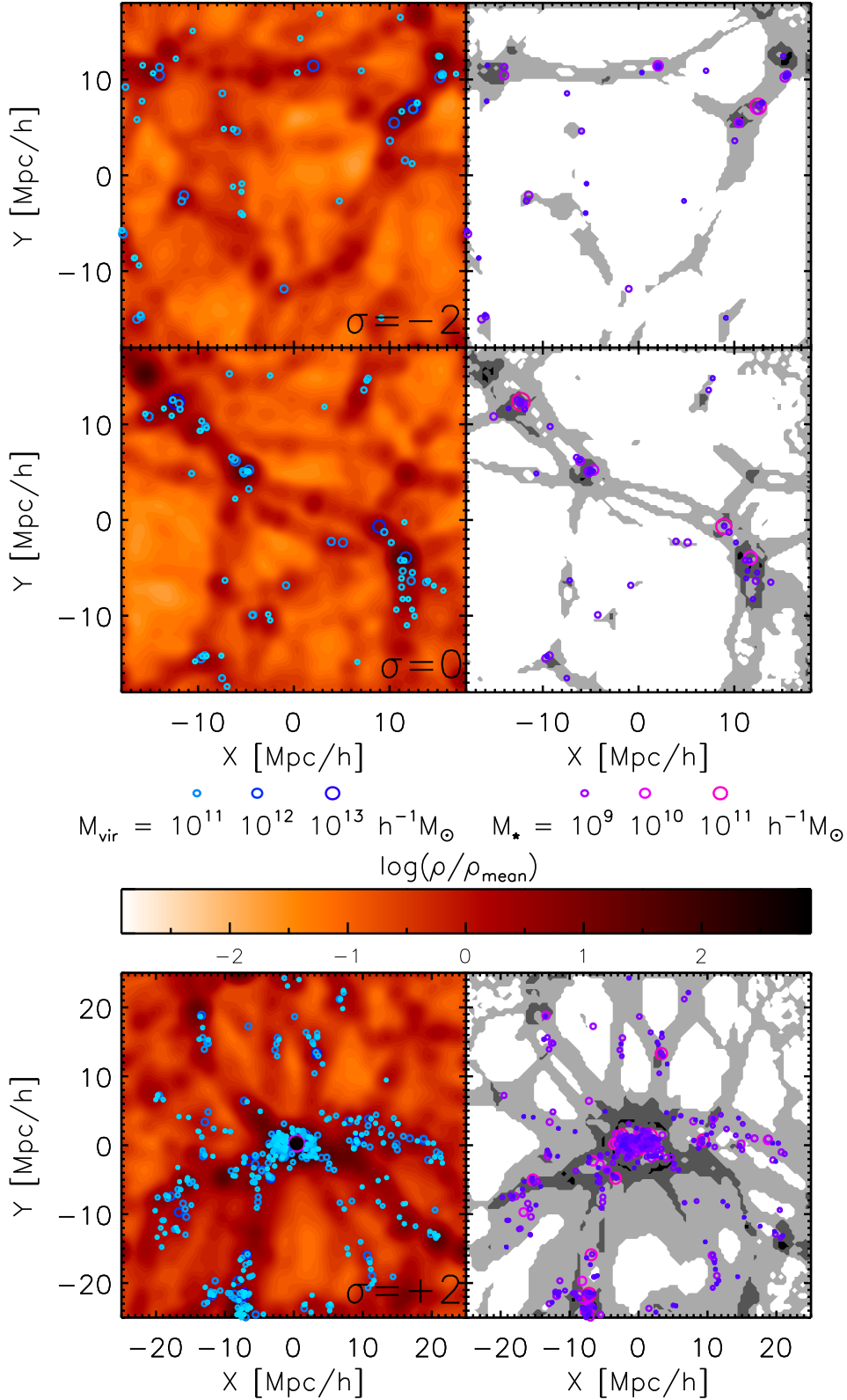}
\caption{Slices of thickness $0.5$h$^{-1}$Mpc of the $\sigma=[-2,0,+2]$ simulations, smoothed with a $390$h$^{-1}$kpc kernel. Left panels - DM parent haloes (blue circles; symbol size is proportional to log$($M$_{vir})$, as indicated by the legend) plotted on top of the logarithm of the DM density. The colour bar describes the logarithm of the density; all densities are scaled by the universal mean density. Right panels - galaxies in the slice (purple-pink circles; symbol size is proportional log$($M$_*)$) plotted on top of the cosmic web. The contours are coloured by web element (white for voids, light grey for sheets, dark grey for filaments and black for knots).} 

\label{fig:web_dens} 
\end{center} 
\end{figure*}

\subsection{Simulations}  
\label{sec:sim}  

The Galaxies-Intergalactic Medium Interaction Calculation (GIMIC) is a gas-dynamical re-simulation of five roughly spherical regions drawn from the Millennium simulation (\citealt{2005Natur.435..629S} - a DM only cosmological simulation in a periodic box of side length $L_{\rm box}=500h^{-1}$Mpc). We give a short description of the simulations, focusing only on the relevant aspects, and refer the interested reader to the thorough explanations provided in  \cite{2009MNRAS.399.1773C}.

The five re-simulated regions were picked to sample different environmental densities: each has a mean overdensity at $z=1.5$ that deviates from the cosmic mean by $[-2, -1, 0, 1, 2]\sigma$, where $\sigma$ is the \textit{rms} mass fluctuation in a sphere of radius $\sim 20h^{-1}$Mpc.

``Zoomed'' initial conditions are constructed and tested to ensure few low resolution (``tidal'') particles end up within the high resolution region. Gas particles are added to the initial conditions and the hydrodynamics are computed using SPH. This is done at various resolution levels for the purpose of convergence testing: at the highest resolution the computational cost prevented the entire suite from being run to $z=0$. As we are interested in the present epoch, we use the intermediate resolution simulations that were run to $z=0$. The mass resolution for these runs is $M_{\rm DM}\sim 5.3 \times 10^7 h^{-1} $M$_{\odot}$ and $M_{\rm gas}\sim 1.16 \times 10^7 h^{-1} $M$_{\odot}$. A fixed co-moving softening length is used for $z>3$. Below this the softening is fixed in physical space at $1 h^{-1}$kpc. 
Convergence tests on the $-2\sigma$ simulation (which has been run to $z=0$ at a higher resolution) demonstrate that for haloes of $M_{vir} \geq 10^{10} $M$_{\odot}$ results are converged \citep{2009MNRAS.399.1773C}, and for that reason we limit ourselves to haloes in that mass range.

The simulations were performed using a modified version of the \texttt{GADGET3} simulation code (last described in \cite{2005MNRAS.364.1105S}), with modules to follow radiative gas cooling and reionization, star formation, kinetic supernova and stellar winds feedback, and chemical abundances. No AGN feedback is included. The same cosmology as employed by the Millennium simulation (which is consistent with WMAP1 parameters) is assumed: $\Omega_m = 0.25$, $\Omega_{\Lambda} = 0.75$, $\Omega_b = 0.0045$, $n_S = 1$, $\sigma_8 = 0.9$, $H_0 = 100\  h\ $km s$^{-1}$Mpc$^{-1}$ where $h=0.73$. Only the $z=0$ snapshot of the five simulations is used in this paper.

\subsubsection{Gas-dynamics, star formation and feedback}

The baryonic physics that GIMIC models includes the following processes:
\begin{itemize}
\item{{\it Gas cooling and photoionisation} were implemented as described in \cite{2009MNRAS.393...99W} - a redshift dependant, spatially-uniform ionizing background is assumed; the cooling rate is computed as a function of redshift, gas density, temperature and composition.}
\item{{\it Quiescent star formation and feedback} were implemented by imposing a polytropic equation of state to gas particles with density $n_H > 10^{-1}$ cm$^{-3}$. These particles are eligible for star formation, and are stochastically converted into star particles with a probability determined by the local gas pressure  \cite{2008MNRAS.383.1210S}.}

\item{{\it Kinetic feedback} was implemented by stochastically imparting randomly oriented velocity kicks to, on average, $\eta = \dot{m}_{\rm wind}/\dot{m}_{\star} = 4$ neighbouring gas particles of newly formed stars. Kicks of 600 km/s were adopted, and implemented $3 \times 10^7$ years after the formation of the star particle. This implementation is described by \cite{2008MNRAS.387.1431D}.}

\item{{\it Chemodynamics} - star particles inherit their abundances in 11 elements (hydrogen, helium, carbon, nitrogen, oxygen, neon, magnesium, silicon, sulphur, calcium, and iron) from their parent gas particles. An initial mass function \citep[][with a stellar mass range of $0.1 - 100 M_{\odot}$]{2003PASP..115..763C} and stellar evolution tracks \citep[dependant on metal abundance; ][]{1998A&A...334..505P,2001A&A...370..194M,2003NuPhA.718..139T} are assumed, and the delayed release of these elements \citep{2009MNRAS.393...99W} from supernovae and AGB stars is then followed as they are being distributed to neighbouring gas particles. Solar abundances are based on CLOUDY.}

\end{itemize}

The lack of AGN feedback and the simplified wind model results in a luminosity function which does not properly match observations at high masses (see Crain et al 2009). However, this mismatch does not affect our main point of interest, namely the relative properties across different environments.

\subsubsection{DM halo and galaxy identification}

DM haloes are identified using the publicly available Amiga Halo Finder \citep[AHF;][]{2009ApJS..182..608K}, which locates local overdensities in an adaptively smoothed density field as prospective halo centres (the global mass function of haloes has been verified to to be consistent with the that derived by \cite{2009MNRAS.399.1773C}, using the SUBFIND algorithm; the `halo finder comparison project' provides a thorough discussion on the comparisons of structure finders of haloes \citep{2011MNRAS.415.2293K,2013MNRAS.435.1618K}, subhaloes \citep{2012MNRAS.423.1200O}, and galaxies \citep{2013MNRAS.428.2039K}). The potential of each density peak is then calculated and bound particles are retained as halo members. AHF implicitly locates both parent and subhaloes, simultaneously (``parents'' are haloes whose centre does not fall within the virial radius of any other halo; subhaloes are haloes contained within the virial radius of another halo). The virial radius is defined in the standard way (namely, the distance at which the mean interior density falls below $200\rho_{\rm back}$, where $\rho_{\rm back}$ is the mean density in the simulation). The radius of a subhalo cannot be defined in this manner since the overdensity may only drop below the given threshold outside the parent halo, and so its extent is defined as the point at which the density profile starts rising, effectively signalling the transition from the subhalo to the main halo.

Following \cite{2012MNRAS.423.1544S} we identify galaxies by examining the particles within 15\% of $r_{\rm halo}$, where $r_{\rm halo}$ is the parent or sub- halo radius. This value is used to ensure that most of the stellar mass in the halo is inside the galaxy \citep[see ][]{2013ApJ...764L..31K}. Within a given halo, the fraction of star particles that reside within $0.15 r_{\rm halo}$ depends on the host halo mass. For example, for high mass haloes ($M_{\rm halo} > 10^{12}h^{-1}$M$_{\odot}$) around 90\% of all halo star particles reside within $0.15 r_{\rm halo}$; this drops to around 50\% for low mass haloes ($M_{\rm halo} \sim 10^{10}h^{-1}$M$_{\odot}$, which is our resolution limit for haloes). In order to ensure our results are not dependent on this choice we have run our analysis using 0.1, 0.15, and 0.2 of $r_{\rm halo}$, and found no differences in our main results, explained in \S\ref{sec:results}. All baryonic particles within this distance define the ``galaxy'' if the number of star particles exceeds 10. Such a galaxy finder yields 19081 central galaxies (with $M_{\rm vir} > 10^{10}h^{-1}$M$_{\odot}$), 733 of which contain satellites. The total number of satellite galaxies in the simulations is 2876. Haloes containing galaxies are defined here as ``luminous haloes''. Groups of galaxies are defined as all galaxies hosted within the same parent halo.

Care needs to be exercised when obtaining global averages from the various co-added GIMIC simulations, owing to the prevalence and size of each region. Thus, when combining results from the different simulations, each region is weighted according to appendix A2 of \cite{2009MNRAS.399.1773C}; specifically haloes and galaxies are added using the weights that appear in table A1 of that appendix.

\subsection{The Web Classification algorithm}  
\label{sec:V-web}  

We use the V-web method \citep{2012MNRAS.425.2049H} to identify the four elements of the cosmic web in the GIMIC simulations, and through this associate haloes and galaxies with their web environment. A cubic sub-volume of side length twice the radius of the re-simulated sphere at $z=1.5$ ($2 \times 25 h^{-1}$Mpc for the $+2 \sigma$ simulation and $2 \times 18 h^{-1}$Mpc for the rest) is identified. The density and velocity fields in this cubic volume are calculated on a $128^3$ grid (see appendix for a discussion of grid sizes and convergence tests) using the Clouds-in-Cells (CIC) algorithm, resulting in a cell size of 390 $h^{-1}$kpc for the $\sigma=+2$ and 280 $h^{-1}$kpc for the rest of the simulations. The CIC has an inherent smoothing on the scale of two cells. However, an additional Gaussian smoothing with a kernel size of $(1,2,3,4)\times$ the cell size of the $\sigma=+2$ simulation is also applied, so as to suppress the anisotropy induced by the Cartesian grid (see the appendix for a discussion of smoothing kernel choice). The reason we apply four different smoothing kernels to the same CIC is explained below. The (dimensionless) velocity shear is defined at each grid cell by,
\begin{equation}
\Sigma_{\alpha \beta} = - \frac{1}{2H_0} \left( \frac{\partial v_{\alpha}}{\partial r_{\beta}} + \frac{\partial v_{\beta}}{\partial r_{\alpha}}  \right)
\label{eq:shear}
\end{equation}
where $\alpha,~\beta=x,~y,~z$ and $H_{0}$ is the Hubble constant. It is calculated by means of Fast Fourier Transform (FFT) over the cubic regions. The effect of the periodic boundary conditions employed by the FFT has been tested against a variety of box sizes and found to be negligible, as is explained in the appendix. The shear tensor is diagonalized at each grid cell. Note a minus sign has been introduced in Eq. (\ref{eq:shear}), so as to have a positive (negative) eigenvalue correspond to a collapse (expansion) along the direction of the corresponding eigenvector.

The cosmic web classification is done by counting the number of eigenvalues of the shear tensor at a particular point in the space (cell) that are above a given threshold. Following \cite{2012MNRAS.425.2049H} and \citet{2012MNRAS.421L.137L} this threshold is taken to be $\lambda_{\rm th}=0.44$, which best matched the density distribution visually. Thus a cell is tagged as a void, sheet, filament or knot if it has 0, 1, 2 or 3 eigenvalues greater than $\lambda_{\rm th}$, respectively.

Haloes (and their galaxies) inherit the web classification from the cell within which their centres are embedded. The smoothing scale (and corresponding velocity field and shear tensor) assigned to each halo is chosen to be the smallest smoothing scale (of the four used) that is greater than a halo's virial radius.

\section{Results}  
\label{sec:results}

\begin{figure*}

\includegraphics[width=17cm]{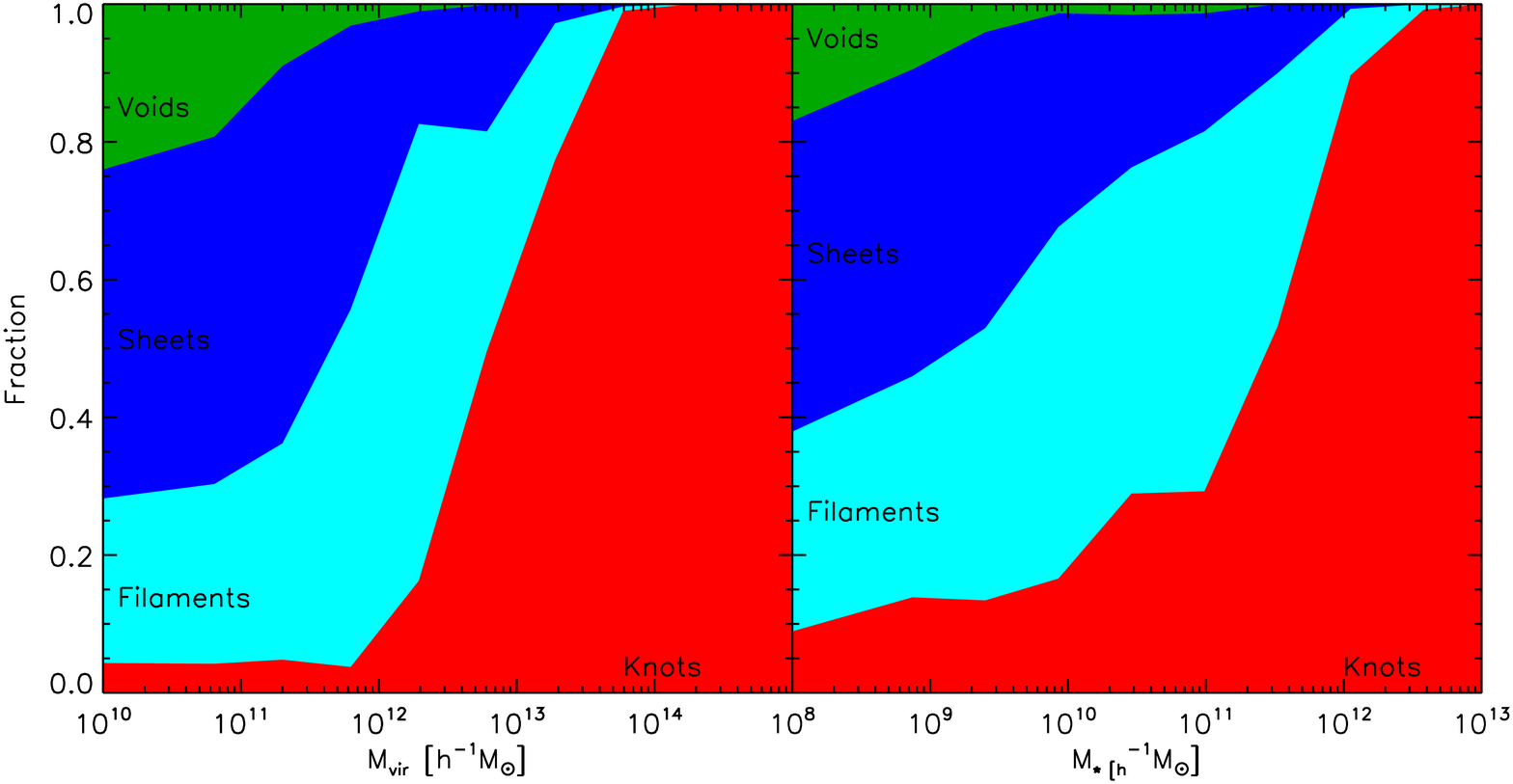}\hspace{0.cm}
\caption{Left: Division of the haloes between the different web elements as a function of halo mass. Right: Division of the galaxies between the different web elements as a function of stellar mass. \\
The coloured regions depict the different elements - red for knots, cyan for filaments, blue for sheets, and green for voids. The fractions are cumulative for each mass; i.e., the upper line of the knots gives the actual fraction of haloes/galaxies in knots at that mass, the upper line of the sheets gives the fraction in knots plus the fraction in sheets, etc.}
\label{fig:hal_gal_frac}
\end{figure*}

The results presented in this section combine properties obtained for each simulation separately, as explained in section \ref{sec:sim}. Care has been taken to ensure that for each web element, these properties do not differ significantly between the five simulations, and so these results are robust across different density environments. 

\begin{table}
  \caption{Statistics of different web elements - their volume filling fractions, the number of haloes and galaxies they host, and the DM and stellar masses they contain, each given as a fraction of the entire simulation. The bottom line gives the fraction of the total haloes (above our resolution limit of $10^{10} $M$_{\odot}$) within each web element that contain galaxies (i.e., luminous haloes).}
  \begin{center}
    \begin{tabular}[l]{l c c c c}
       & Voids & Sheets & Filaments & Knots \\
      \hline
      Volume fraction & 63.7\% & 31.7\% & 4.3\% & 0.3\% \\
      Halo no. fraction & 15.9\% & 51.5\% & 28.8\% & 3.8\% \\
	  DM mass fraction& 16.2\% & 35.4\% & 33.0\% & 15.4\% \\
	  Galaxy no. fraction & 13.4\% & 46.5\% & 30.8\% & 9.3\% \\
	  Stellar mass fraction & 0.8\% & 11.3\% & 36.4\% & 51.5\% \\
	\hline
	  Fraction of haloes & 49.0\% & 63.2\% & 75.0\% & 80.2\% \\
	  that are luminous & & & & \\
      \hline
    \end{tabular} 
  \end{center}
  \label{table:percent}
\end{table}

In Fig. \ref{fig:web_dens} we show a $0.5 h^{-1}$Mpc thick slice through the centre of the $[-2,0,+2] \sigma$ GIMIC regions, smoothed with a $390$h$^{-1}$kpc smoothing kernel. On the left we show the logarithm of the density, with DM haloes over-plotted (blue circles; circle radius scales with the radius of the halo), while on the right, we show the cosmic web, decomposed into voids, sheets, filaments and knots with the galaxy distribution over-plotted (purple-pink circles; the radius of the circles is proportional to the logarithm of the galaxy's stellar mass). That haloes and galaxies trace the matter distribution is expected and clearly seen in these plots.

\begin{figure*}

\includegraphics[width=8.5cm]{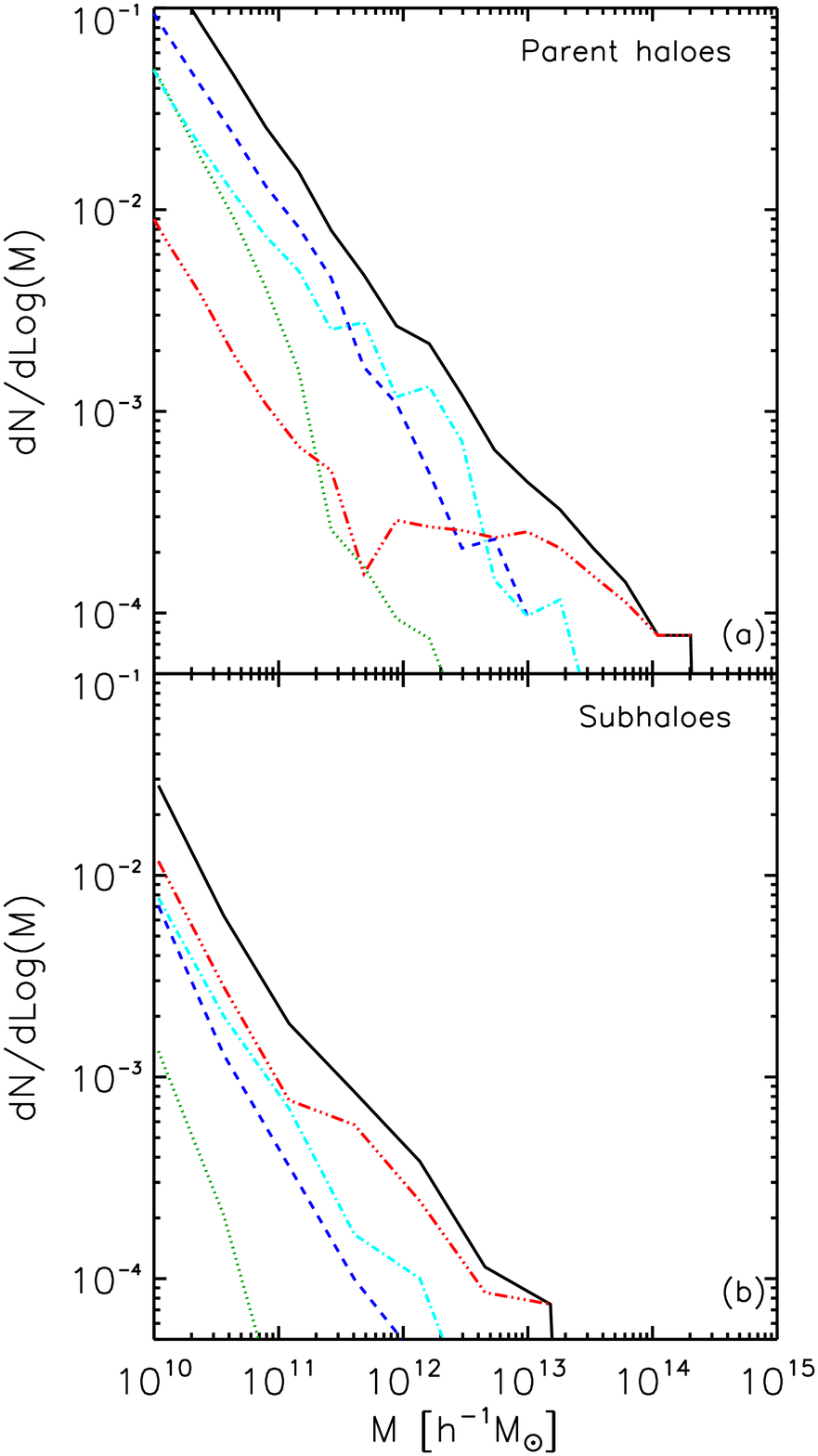}
\includegraphics[width=8.5cm]{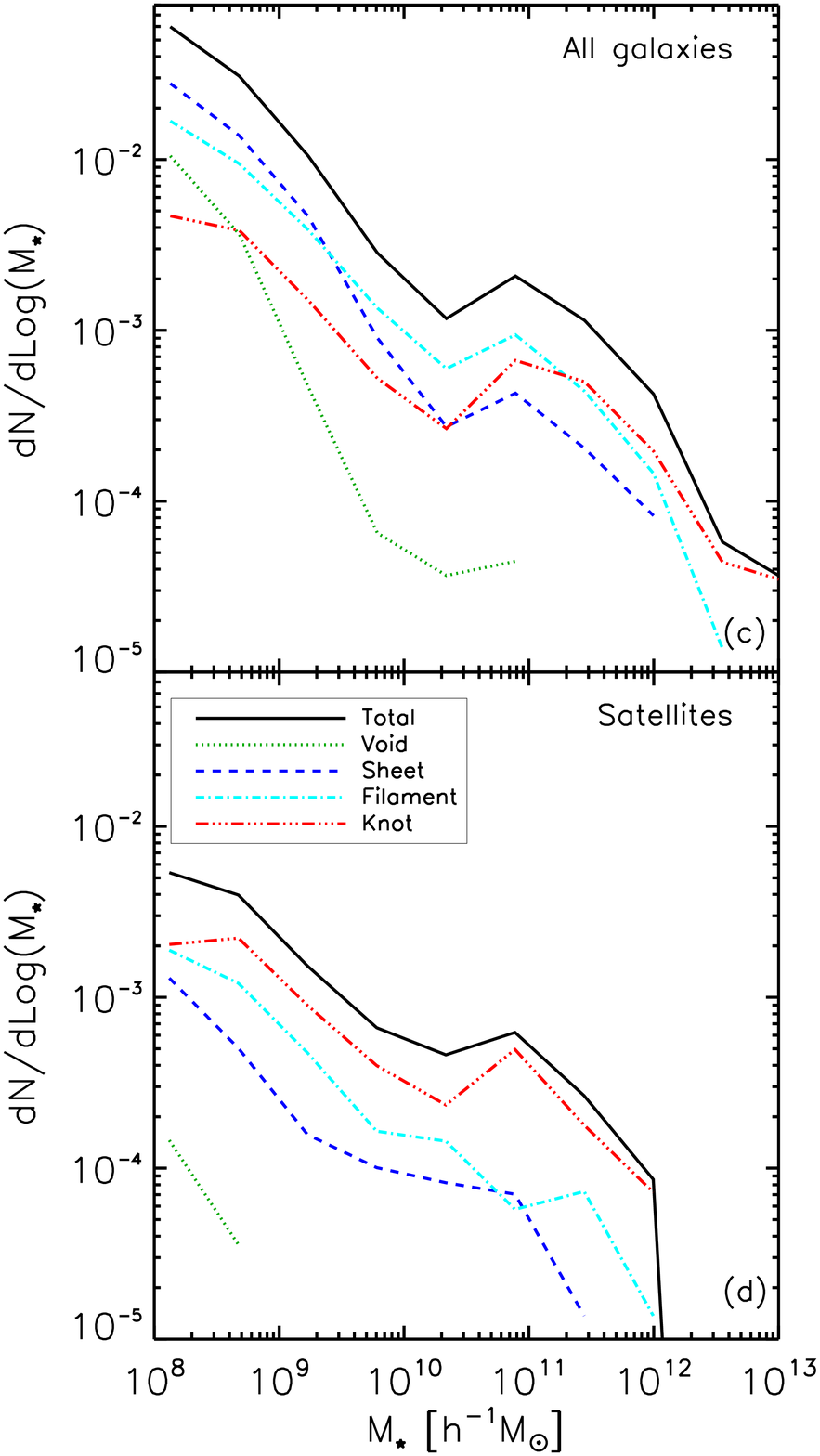}

\caption{Left: The DM mass functions of all (top) and sub- (bottom) haloes. Right: The stellar mass functions of all (top) and satellite (bottom) galaxies. \\
  Different line styles denote the total (solid black line), voids (green dotted line), sheets (blue dashed line), filaments (single dot-dashed cyan line), and  knots (triple dot-dashed red line).}
\label{fig:mass_lum_funcs} 
\end{figure*}

The statistics which characterize the cosmic web are presented in Table \ref{table:percent}. It includes the volume fractions, the DM and stellar mass fractions, and the fractions of haloes and galaxies in each web type, as well as the fractions of haloes above our resolution limit that contain galaxies in each web type (defined here as luminous haloes, as opposed to dark haloes that contain no luminous galaxies; for an examination of dark haloes in the GIMIC simulations see \cite{2013MNRAS.431.1366S}). The following conclusions follow. Haloes, galaxies and the DM reside predominantly in filaments and sheets. Voids dominate by volume, with a volume fraction of $\sim64\%$, yet contain less than 1\% of the stellar mass. These results agree with works using different web classification schemes, such as \cite{2007A&A...474..315A} and \cite{2014MNRAS.441.2923C}.
The number fraction of galaxies distributed by their web classification constitutes a fairly close proxy to that of haloes and to the DM mass fraction. This is not the case for the stellar mass whose web distribution varies significantly from that of the DM. 

The fact that knots contain roughly half of all stellar mass available is a reflection of the following two properties. 
The first is inferred from the bottom row of Tab. \ref{table:percent}, which shows the fraction of DM haloes above our resolution limit in each web type that host galaxies (``luminous haloes''). Approximately $\sim80\%$ of haloes in knots have luminous galaxies, whereas less than a half of haloes in voids have luminous galaxies. The second is the different mass functions \footnote{Throughout this paper, we use the term "mass functions", plural, to denote the division into the four web elements.} (see Fig~\ref{fig:mass_lum_funcs}).

The left (right) panel of Fig. \ref{fig:hal_gal_frac} manifests this fact, by showing the fraction of haloes (galaxies) in each web element as a function of M$_{vir}$ (M$_*$). At low masses, most haloes and galaxies reside in sheets, at intermediate masses most of them are in filaments, and at high masses practically all of them reside in knots. This is in good agreement with the bottom panel of Fig. 18 of \cite{2014MNRAS.441.2923C}.

\begin{figure*}
\includegraphics[width=17cm]{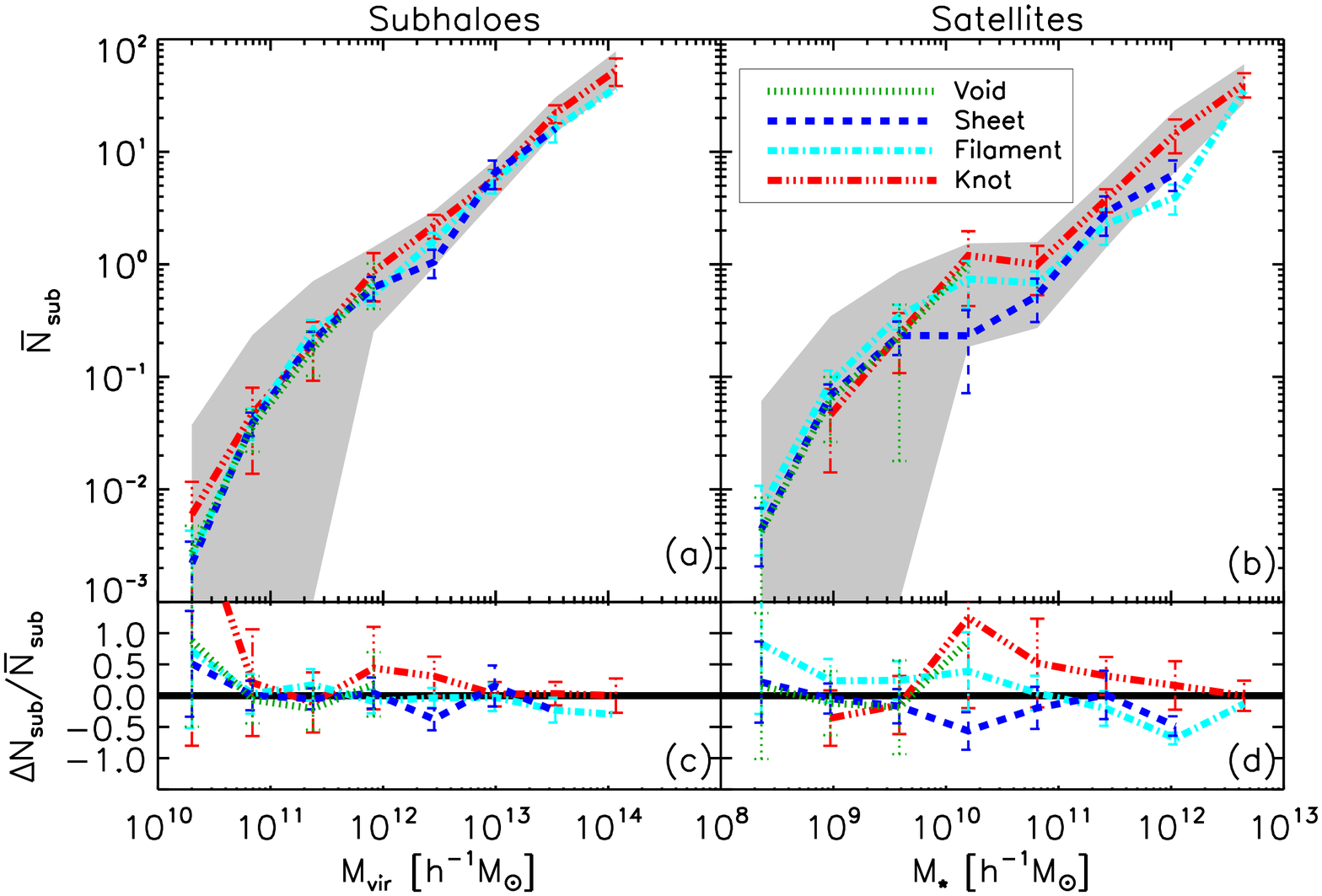}\hspace{0.cm}
\caption{(a) Average number of subhaloes as a function of host halo virial mass. The shaded region depicts the standard deviation of the mean, the error bars depict the standard error of each web element. (b) Average number of satellite galaxies as a function of central galaxy stellar mass. The shaded region and error bars are as in (a). (c) The fractional difference from the mean number of subhaloes for each web element. (d) The fractional difference from the mean number of satellites for each web element.
The average substructure numbers don't change significantly between web elements.}
\label{fig:sub_sat}
\end{figure*}

It is important to understand how the different web classification selects haloes of different DM mass, galaxies of different stellar mass or luminosity, and so on. A number of these properties are shown in Fig.~\ref{fig:mass_lum_funcs}(a-d).

In Fig.~\ref{fig:mass_lum_funcs}(a) we show the DM mass functions of parent haloes. The mass functions vary across the cosmic web, both in amplitude, shape and (importantly) range. Sheets contain most of the lowest mass parent haloes, up to $3\times10^{11}h^{-1}$M$_{\odot}$. Most haloes of mass between $3\times10^{11}h^{-1}$M$_{\odot}$ and $10^{13}h^{-1}$M$_{\odot}$ are found in filaments, while the most massive haloes in the simulation ($M_{vir}>10^{13}h^{-1}$M$_{\odot}$) are found almost exclusively in knots. Voids contain the smallest fraction of haloes with DM masses above  $2\times10^{11}h^{-1}$M$_{\odot}$. The general trend is similar to the ones seen in \cite{2007A&A...474..315A} and \cite{2014MNRAS.441.2923C}, despite the fact that the web classification schemes used in those papers are different; this is not entirely surprising, as the most massive haloes are expected to reside in knots, while voids are expected to be mostly empty and definitely not contain any massive galaxies, and so on.

The subhalo mass function is shown in Fig.~\ref{fig:mass_lum_funcs}(b). From this figure we infer that most subhaloes are embedded in knot-parents and the fewest subhaloes are embedded in void haloes (irrespective of subhalo mass). This is likely due to the fact that the parent knot halo mass function extends to higher masses and thus, above the given resolution limit, contain more substructures than all of the other web elements.

The right panels of Fig.~\ref{fig:mass_lum_funcs} show various galaxy stellar mass functions. In Fig.~\ref{fig:mass_lum_funcs}(c) the full galaxy stellar mass function is shown. It is similar to the halo mass function in the fact that knot environments host the most massive galaxies (where $M_{\star}\gsim10^{12}h^{-1}$M$_{\odot}$), with filaments and sheets dominating the environments of the lower mass bins ($10^{10}\lsim M_{\star}/h^{-1}$M$_{\odot}\lsim10^{12}$ and $M_{\star}\lsim10^{10}h^{-1}$M$_{\odot}$, respectively).  Note that the apparent "bump" at stellar masses of $\gsim 10^{11} $M$_{\odot}$ is driven by the lack of AGN feedback and the simplified wind model in the GIMIC subgrid physics. It is also interesting to note that in the lower mass end, all total mass functions (both DM and stellar) follow a power law with the same slope of -1.1, at least up to the point where the ``bump'' in the stellar mass function begins.

The satellite stellar mass function, Fig.~\ref{fig:mass_lum_funcs}(d), is also broadly similar to its DM counterpart in showing a clear hierarchical trend for the stellar mass functions of the different web elements across the entire range of stellar masses. However, practically no satellite galaxies above our resolution limit inhabit void regions - this is likely because DM haloes in voids are on average less massive and thus have fewer resolvable substructures. This limits the number of both substructures and satellites.

This comparison between the left column of Figure \ref{fig:mass_lum_funcs}, which describes the mass function of the DM haloes, with the right column, which shows the various stellar mass functions, and their dependence on the web environment, leads to the inference that galaxy-environment dependence (right column) is strongly driven by the halo mass-environment dependence (left column). This seems even likelier when considering the fact that the trends do not change whether one plots the stellar mass functions of all haloes, groups of galaxies, or just the centrals. \\

In Fig.~\ref{fig:sub_sat} we show the average number of subhaloes per host halo (a) and satellites per host galaxy (b) as a function of host mass, divided into web types. The shaded regions depict the standard deviation about the mean for all haloes/galaxies, while the error bars depict the standard error of the mean (i.e., the error involved in computing the mean; these errors were computed using a bootstrapping algorithm) about the mean for each web element. The lower panels show the fractional difference from the mean, $\Delta(N_{sub})/ \overline{N}_{sub}$, for the different web elements.
As can be seen, while there is some difference in the number of satellite galaxies at a given galaxy stellar mass, it is small and within the standard error of the mean for almost all masses and web elements. For the number of subhalos at a given halo mass the difference is even smaller. The curves are almost entirely within the standard deviation around the mean. \\

\begin{figure}
\begin{center} 
\includegraphics[width=8cm]{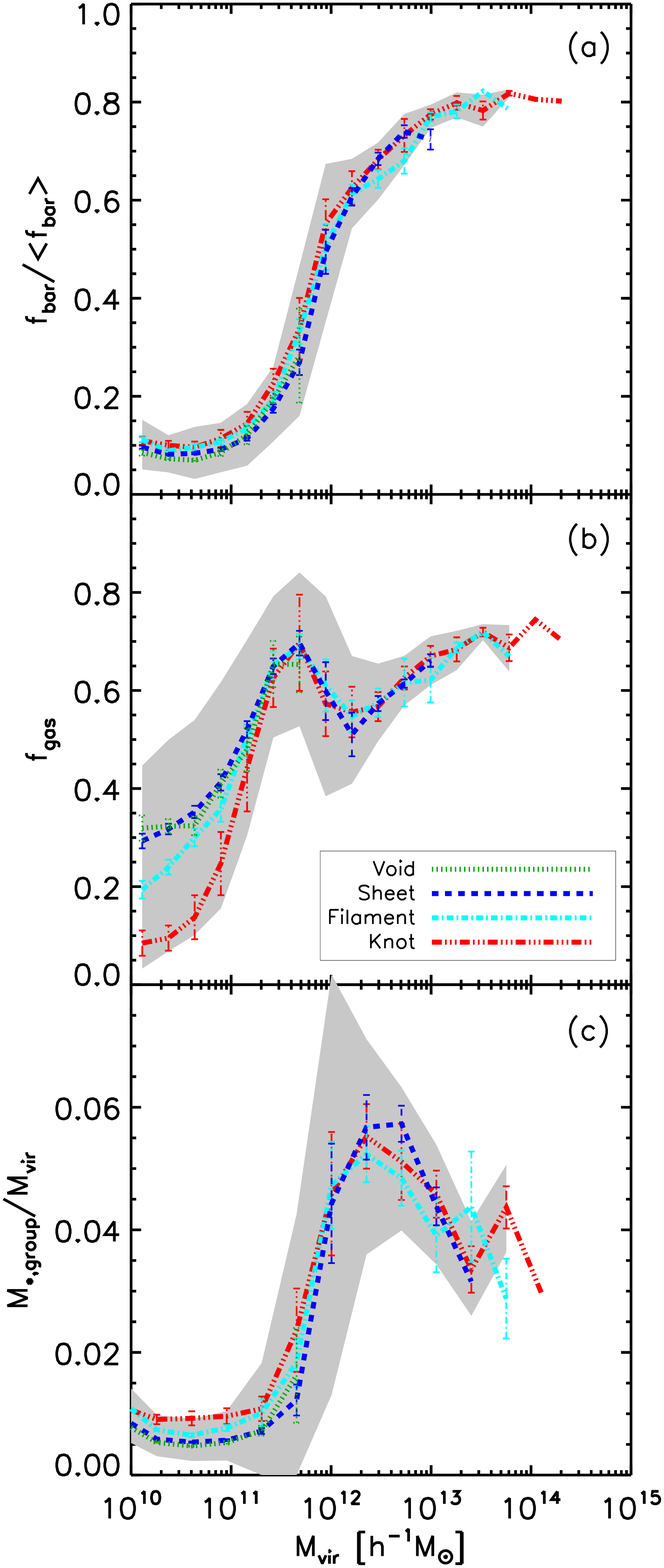}
\caption{Three properties of the baryonic component of luminous haloes as a function of the halo's virial mass. Panel (a) shows the baryon mass fraction (normalized by the mean baryon mass fraction in the simulation); panel (b) presents the gas fraction of the baryonic component; panel (c) gives the total stellar mass in galaxies within the halo divided by the halo's virial mass. The shaded are depicts the standard deviation around the mean for all haloes, and the error bars are the standard error of  the mean for each web element. All three plots show a distinctly small difference between the different web elements, with the exception of gas fractions at small host halo masses. All of the curves are within the $1 \sigma$ range around the mean.}
\label{fig:fbar_fgas} 
\end{center} 
\end{figure}

In Fig.~\ref{fig:fbar_fgas} we examine three properties of the baryonic component in luminous haloes as a function of the host halo's virial mass. The shaded regions depict the standard deviation around the mean for all haloes, while the error bars show the standard error of the mean for each web element.

In Fig.~\ref{fig:fbar_fgas}(a) we show the mean baryon mass fraction in haloes ($f_{bar}=M_{bar}/M_{vir}$), divided by the average baryon fraction in the simulation. The baryon mass fraction increases with host halo mass, in agreement with previous works \citep{2008MNRAS.391..481S,2010ApJ...708L..14M}, but the differences between web elements for a given halo mass are very small. The baryon mass fraction never reaches the universal mean, in agreement with \cite{2007MNRAS.377...41C}.

In Fig.~\ref{fig:fbar_fgas}(b) we show $f_{gas}$ - the gas baryon mass fraction (i.e., $M_{gas}/M_{bar}$). The gas fraction increases with halo mass for haloes in all web types. The increase with halo mass is in agreement with the recent work of \cite{2014arXiv1401.0737N}. At low halo masses ($< 10^{11}h^{-1}$M$_{\odot}$) the gas fractions of the different web elements differ visibly, though still within a factor of a few. All curves lie within the standard deviation around the mean. In this mass range knot haloes have a smaller gas fraction than filament haloes, which in turn have a smaller gas fraction than sheet and void haloes. This may be due to the low baryon masses at these host halo masses - as can be seen from \ref{fig:fbar_fgas}(a), at these halo masses the baryon fraction is much lower than for higher halo masses, and so a slight change in star formation can mean a significant change in gas fraction.

In Fig.~\ref{fig:fbar_fgas}(c) we show $M_{\star,group}/M_{vir}$, the fraction of the virial mass that is locked up in galaxies. For low mass haloes, the stellar mass component increases with host halo mass, and there is some difference between the different web elements, though within a factor of 2. A transitional mass between $10^{12}h^{-1}$M$_{\odot}$ and $10^{13}h^{-1}$M$_{\odot}$ exists, above which the fraction of halo mass in stars starts to decrease with host halo mass. This behaviour is in agreement with previous works \citep{2008ApJ...676..248Y, 2010ApJ...718.1001B, 2010ApJ...708L..14M}. These trends are irrespective of web type, and all curves are within the standard deviation around the mean.

Similar trends are visible across web types in all three plots, indicating again that, to first order, haloes behave similarly irrespective of web environment. Although the overall shapes of these plots agree with previous works, we do not stress the shapes of these plots too much, as they may be affected by the physics of the simulation. However, this should not change the underlying result - the lack of difference between the web elements for a given halo mass (with the exception of a small difference at small halo masses for some properties).\\

\begin{figure} 
\begin{center} 
\includegraphics[width=8cm]{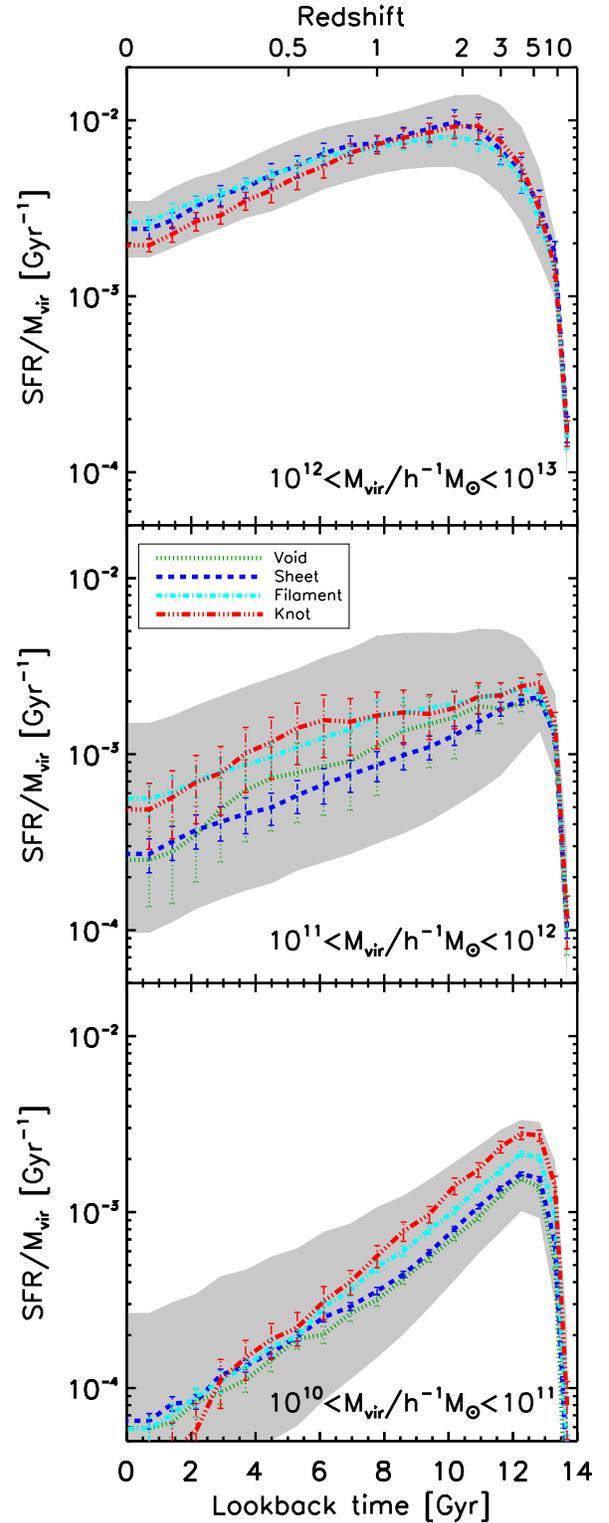}\hspace{0.cm}
\caption{The star formation rate in central galaxies normalized by host halo mass as a function of lookback time in three different host halo mass bins (in units of $h^{-1}$M$_{\odot}$), divided into web elements. Shaded regions depict the standard deviation around the mean for all galaxies, while error bars show the standard error of the mean for each web element. While the behaviour changes between different mass bins, there is very little change between web elements in each mass bin; even at small masses, the plots lie outside of the standard errors of each other at early times, the change is within a factor of 2. All curves lie well within the standard deviation around the mean.}
\label{fig:sSFR} 
\end{center} 
\end{figure}

Is the ability to turn baryons into stars affected by the web environment? We investigate this possibility in Fig.~\ref{fig:sSFR}, which shows the star formation rate (SFR) divided by host halo mass as a function of lookback time for three host halo mass bins, and divided into web elements. Effectively, we calculate a (differential) histogram of ages for star particles that are bound to galaxies at z = 0. The SFR is the total mass of all star particles formed during a finite time interval divided by that time interval. The shaded regions depict the standard deviation around the mean for all galaxies, while the error bars show the standard error of the mean for each web element.

As can be seen, haloes of different masses have different star formation histories, the main difference being that high mass haloes have much higher star formation at later times than intermediate and low mass haloes, with the peak of star formation between redshift 1 and 2. We do not stress the shapes of these plots, as they have been shown to be difficult to reconcile with observations \citep{2012MNRAS.426.2797W}. However, it is also seen that for high mass haloes, the difference between the different web elements is very small at all times. Intermediate mass haloes show slightly more difference for later epochs, but still within a factor of a few at most. Low mass haloes show distinct differences at early times, and especially at the peak, but these are small (within a factor of 2). At all masses the curves lie well within the standard deviation around the mean. The fact that the differences are small appears to be robust, and independent of the shapes of the plots. Therefore, in agreement with \cite{2009MNRAS.399.1773C}, we find that the main driver of the different SFRs is driven by its dependence on host halo mass, and through it an environmental dependence is induced. 
 
\section{Discussion}  
\label{sec:disc}  

One of the very basic components of any theory of galaxy formation is the dependence of at least some of the properties of galaxies on their ambient environment. Arguably, the morphology-density relation of  \cite{1980ApJ...236..351D} is the best studied manifestation of such a dependence. This has led us to analyse the GIMIC set of simulations \citep{2009MNRAS.399.1773C}, a suite of state of the art gas-dynamical simulations that include a myriad of physical phenomena. It is convenient to define the environment within which galaxies reside by the cosmic web, which consists of voids, sheets, filaments and knots  \citep{1996Natur.380..603B}. The velocity shear tensor formulation \citep{2012MNRAS.425.2049H, 2012MNRAS.421L.137L} has been used here to quantify the cosmic web. Our analysis consists of studying the properties and distribution of simulated DM haloes and luminous galaxies with respect to the cosmic web. 

The main results of the paper are: 
\begin{itemize}
\item The DM halo mass function varies significantly with the web classification, namely the more massive a halo is the more likely it is to reside higher in the web sequence (voids, sheets, filaments and knots). A similar trend is found when the sample of haloes is limited only to luminous ones, namely the ones who contain a galaxy.

\item The stellar mass functions of all galaxies vary significantly with the web elements, in a manner similar to the halo mass functions. Voids are populated mostly by faint galaxies, and moving up along the web sequence, sheets, filaments and knots are progressively populated by more massive galaxies. A similar trend is found for the stellar mass functions of groups of galaxies and of central galaxies, where the group is defined by all galaxies embedded in a given parent halo.

\item Subhaloes show a more distinct trend with the web sequence, showing dominance according to web sequence in all subhalo masses. Satellite galaxies have a similar trend with stellar mass.

\item The number of satellites per galaxy of a given mass extends over more than three orders of magnitude for the range of haloes studied here, with a relatively small dependence on the web environment. The number of subhaloes per halo of a given mass shows an even smaller dependence on web environment.

\item Considering the properties of baryons residing in DM haloes at a given mass range, these properties are broadly independent of the web classification. The baryon mass fraction, gas fraction, and stellar mass fraction in luminous haloes depend strongly on the virial mass of the halo, and show very little direct dependence on the web environment for most of the mass range studies. Gas fractions at low halo masses do seem to differ noticeably between web elements; however, this is still within a factor of a few, and may be a result of the low baryon masses in these haloes. All curves lie within the standard deviation around the mean.

\item The star formation rate of central galaxies, divided by host halo mass, shows slight variation with respect to web type. For high mass haloes, the different web types are practically indistinguishable within the errors; intermediate mass haloes show some variation, but still within the standard errors; low mass haloes show the most significant changes, but these are still small, a factor of 2 at the most. All curves lie well within the standard deviation around the mean. In agreement with the other 'baryonic' properties of galaxies residing in DM haloes, the difference seen between different mass bins is induced by the strong dependence of the star formation on the host halo virial mass.

\end{itemize}

The following understanding of the galaxy-environment relation emerges. There is a clear and strong dependence of the halo mass function on the web classification. It has been long known that denser environments host more massive haloes, but this dependence has been quantified now in terms of the cosmic web classification that is adopted here. Galaxies are characterized here by a few global properties, such as gas and stellar mass fraction, stellar mass and star formation. The distribution of haloes and galaxies with respect to these properties depends on the web type and this is best manifested by the dependence of the galaxy stellar mass function on the cosmic web attributes. Again, a very well known fact but quantified here within the framework of the cosmic web. Looking deeper into the web dependence the distribution of the above properties of haloes and the galaxies that they host, within a given mass range, has been studied, again in the context of the cosmic web. 

Our main finding is that as a first approximation the dependence of baryonic properties of galaxies on their ambient web environment is driven by the different mass function of the different web environments. One should note that it is not claimed here that environmental effects do not shape the properties of galaxies. Rather, to the extent that environment shapes the halo mass function, it also shapes the galaxy properties, but indirectly. The end product of such effects is that the halo mass is the main parameter through which the web environment affects processes related to galaxy formation.

The major conclusion of this paper is that the strong dependence of galaxy properties on the mass of their host halo is the leading effect that shapes the dependence of the galaxy properties on the cosmic web. For the most part, the baryonic parameters are indistinguishable within the standard errors, with the exception of gas fractions, stellar masses, and SFR/M$_{vir}$ of low mass haloes ($10^{10}-10^{11}$h$^{-1}M_{\odot}$). Even these are within a factor of a few, and we can therefore conclude that for the GIMIC simulation the web environment affects the properties of galaxies only to a small degree, and that only for low mass haloes.

This lends support to a wide range of semi-analytical models \citep[e.g.][]{2000MNRAS.319..168C,2006MNRAS.370..645B} and statistical schemes of populating haloes with galaxies \citep[e.g. CLF;][]{2007MNRAS.376..841V}, all of which assume that the galaxy properties are determined by the mass of a halo and its mass aggregation history. \\

\begin{center}
\textbf{Acknowledgements}
\end{center}

NIL is supported by the Deutsche Forschungs Gemeinschaft

YH has been partially supported by the Israel Science Foundation (1013/12).

This work was supported by the Science and Technology Facilities Council [grant number ST/F001166/1], by the Interuniversity Attraction Poles Programme initiated by the Belgian Science Policy Office ([AP P7/08 CHARM], and used the DiRAC Data Centric system at Durham University, operated by the Institute for Computational Cosmology on behalf of the STFC DiRAC HPC Facility (www.dirac.ac.uk). This equipment was funded by BIS National E-infrastructure capital grant ST/K00042X/1, STFC capital grant ST/H008519/1, and STFC DiRAC Operations grant ST/K003267/1 and Durham University. DiRAC is part of the National E-Infrastructure. The data used in the work is available through collaboration with the authors. \\

\bibliography{./ref}

\appendix

\section{Appendix: Numerical convergence tests}
\label{sec:appendix}

The web formalism includes a number of steps that could potentially alter our results. We have taken care to ensure that this is not the case.

\subsection*{Periodic boundary conditions}

The V-web algorithm uses a the Fast Fourier Transform (FFT) to calculate the spatial derivatives of the velocity field. The application of the FFT assumes periodic boundary conditions for the computational box. This is not the case in the present analysis of the GIMIC zoom sub-boxes and this might lead to some numerical artefacts in a 'boundary layer' that is a few cells thick (as was shown in \cite{2012MNRAS.425.2049H}). However, this has virtually no effect in the present case where the CIC box is chosen to enclose the almost spherical zoom region, and haloes are chosen only from that spherical region. Hence only very few high-resolution cells are affected.

\subsection*{Smoothing kernel size}

\begin{figure}
\begin{center} 
\includegraphics[width=8cm]{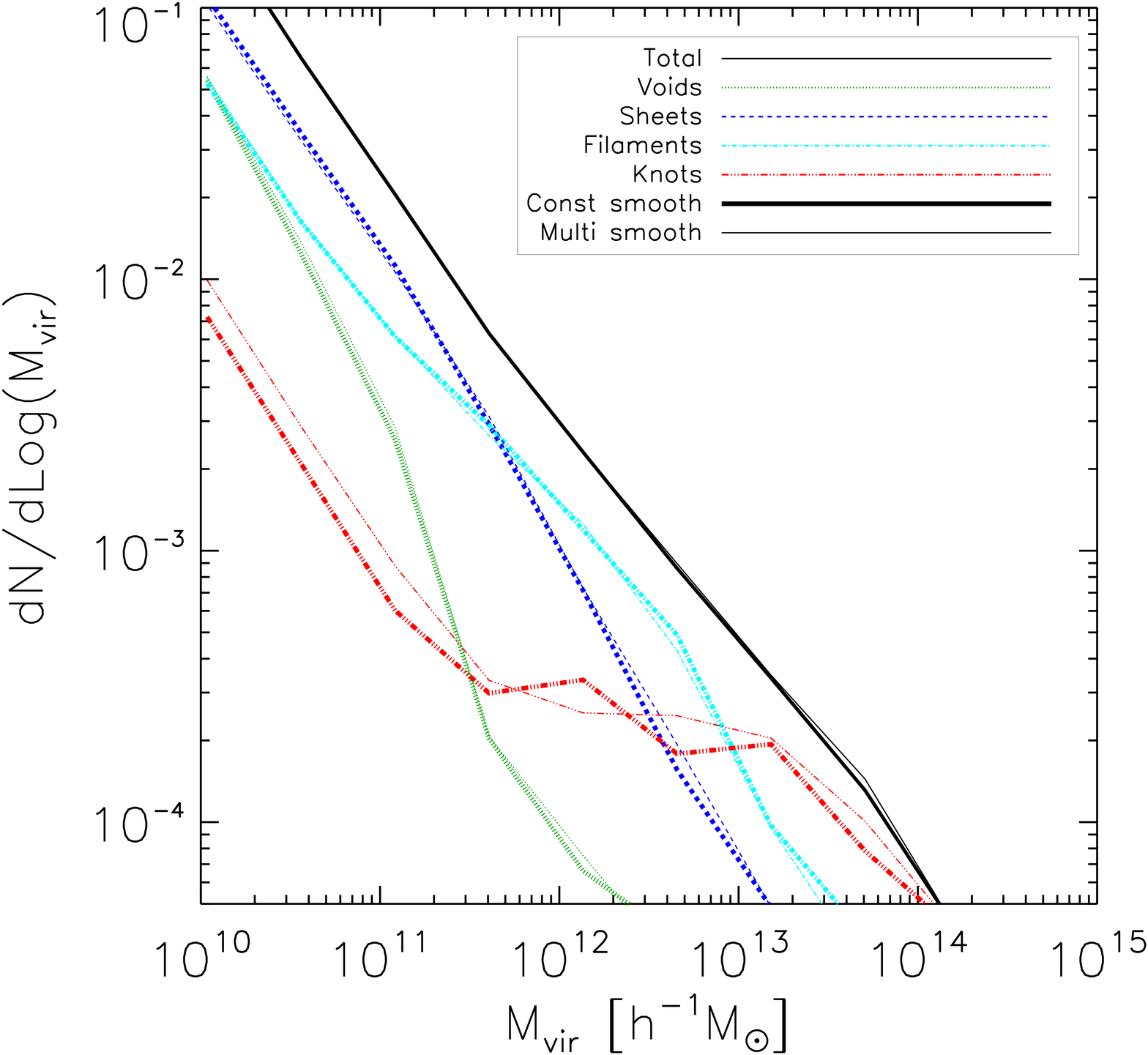}
\caption{The mass functions of parent haloes, comparing the case of a constant smoothing kernel size (Const smooth; solid lines), as was used in this work, and of a smoothing kernel size that changes according to the size of the computational box (Multi smooth; dashed lines). As can be seen, the two smoothing kernel sizes produce very similar curves.}
\label{fig:mass_func_compare} 
\end{center} 
\end{figure}

That the cell sizes are different between the different simulations is unavoidable since the volumes of their area of interest (the high resolution spheres) are different, and the FFT algorithm is fastest when the grid size is a factor of 2. In order to be able to compare the different simulations, the same smoothing kernel sizes (1-4 cell sizes of the $+2\sigma$ simulation) were used in all the simulations. To ensure that this choice does not affect our results we have made a similar analysis with smoothing kernel sizes that are 1-4 cell lengths for each simulation, rather than of the +2$\sigma$. All our results are reproduced in that case; as an example we bring a comparison of the mass functions of parent haloes, shown in Fig. \ref{fig:mass_func_compare}.

\subsection*{Grid size}

In this work we use the intermediate resolution set (MidRes) of GIMIC simulations since not all simulations in the high resolution set (HiRes) have been run to redshift zero. This forces us to use a $128^3$ grid in order to ensure that there is (on average) at least one particle per cell. However, in order to ensure that the choice of grid size does not affect our results, we have performed resolution tests on different sized grids; since for the MidRes simulation a $256^3$ grid is too sparse, we have used the HiRes version of the $-2\sigma$ simulation, which has been run to redshift zero, and compared it also to the MidRes simulation. This was done by comparing the volume fractions of the different web elements from the webs computed on three grids - a $128^3$ grid in the MidRes simulation, a $128^3$ grid in the HiRes simulation, and a $256^3$ grid in the HiRes simulation, all with the same smoothing kernel size. The results are summarized in Table \ref{table:vol_fracs}, and as can be seen, they are extremely similar for all grids.

\begin{table}
  \caption{Convergence test for the grid size. Shown are the volume fractions occupied by each web element in the $-2\sigma$ simulation, for three different griddings - a $128^3$ grid from the MidRes simulation, a $128^3$ grid from the HiRes simulation, and a $256^3$ grid from the HiRes simulation. }
  \begin{center}
    \begin{tabular}[l]{l c c c c}
       & Voids & Sheets & Filaments & Knots \\
      \hline
      $128^3$ MidRes & 76.9\% & 20.5\% & 2.5\% & 0.1\% \\
      $128^3$ HiRes & 78.4\% & 19.3\% & 2.2\% & 0.1\% \\
	  $256^3$ HiRes & 76.6\% & 20.9\% & 2.4\% & 0.1\% \\
	  \hline
    \end{tabular} 
  \end{center}
  \label{table:vol_fracs}
\end{table}

\end{document}